\documentclass[a4paper,twoside]{article}

\usepackage{epsfig}
\usepackage{subcaption}
\usepackage{calc}
\usepackage{amssymb}
\usepackage{amstext}
\usepackage{amsmath}
\usepackage{amsthm}
\usepackage{multicol}
\usepackage{pslatex}
\usepackage{apalike}
\usepackage{todonotes}
\usepackage{algorithm2e}
\usepackage{xcolor}
\usepackage{array}
\usepackage{tabularx}
\usepackage{hyperref}
\usepackage{multirow}
\usepackage{makecell}
\usepackage{booktabs}
\usepackage[bottom]{footmisc}
\usepackage{SCITEPRESS}     

\begin{document}



\title{A Pilot Study on Detecting Software Design Patterns with Large Language Models: An Empirical Evaluation}

\author{
\authorname{Oishik Chowdhury, Bastin Tony Roy Savarimuthu and Sherlock A. Licorish}
\affiliation{School of Computing, University of Otago, New Zealand}
\email{(oishik.chowdhury@postgrad.otago.ac.nz, tony.savarimuthu@otago.ac.nz, sherlock.licorish@otago.ac.nz)}
}



\abstract{Design patterns provide reusable solutions to recurring software design problems. Automatically detecting these patterns in source code can help bootstrap new developers’ understanding of unfamiliar software system architectures, and can help experienced developers to quickly identify and rectify potential quality issues. While many prior research works have explored graph based and machine-learning based detection techniques, this work evaluates the design pattern recognition capabilities of four Large Language Models and two ensemble approaches consisting three out of the four models. We also compare their performance when prompted with a) Source code, b) PlantUML representation of source code, and c) Text-based descriptions of the source code. We investigate the detection of five design patterns: singleton, adapter, bridge, composite and decorator. Our preliminary results indicate that LLMs show promise for automatically detecting design patterns, with NextCoder and Gemma 3 demonstrating comparatively higher accuracy than other models evaluated, and the ensemble approaches enhancing the overall efficiency of design pattern detection. We identify several directions for future work.}

\onecolumn \maketitle \normalsize \setcounter{footnote}{0} \vfill

\section{INTRODUCTION\footnote{\color{red}{This paper will appear in the proceedings of ENASE 2026.}}}
\label{sec:intro}

Design patterns are reusable solutions to commonly occurring problems during software design and development \cite{gamma1995design}. A design pattern functions as a template or blueprint for solving these problems, allowing a developer to adapt the pattern to create a working solution rather than just replicating or reusing existing code. This introduces design flexibility and adaptability that static, pre‑written code cannot provide, allowing the underlying structure to be tailored, extended, or dynamically adapted to meet evolving requirements \cite{niculescu2021mixin}. When relying on the reuse of pre-written code, a developer is restricted to its existing implementation, which may not align with the specific needs or constraints of the software being developed. Using design patterns, thus, provide  the opportunity for practitioners to adopt reusable solutions, increasing the maintainability and readability of code \cite{alghamdi2014impact}. However, to realize these benefits, design patterns must be implemented correctly and applied where needed (at the correct location in the project's code hierarchy). Incorrect implementation of design patterns reduces the quality of code, potentially introducing anti-patterns and code smells into the code \cite{cardoso2015co}. Anti-patterns are commonly used solutions in software system design that look helpful, but actually end up causing issues in codes. A catalog of more than 30 such anti-patterns have been defined by researchers (\cite{brown1998antipatterns} and \cite{fowler2018refactoring}). On the other hand, code smells are characteristics of source code that correspond to an underlying design problem, which indicates a need for code refactoring which can be costly. Therefore, detecting the incorrect design patterns becomes a real challenge as it is pursued manually, highlighting the need for automated approaches for design pattern detection and reparation. In order to leverage the recent developments in Large Language Models as tools in assisting the automated detection and reparation of the design patterns by developers, it is essential to understand the capabilities and limitations of LLM-based detection. Subsequently, models can be used to generate corrections or advise on where and how to use patterns correctly for accruing benefits. This position paper tackles the first step in leveraging LLMs for aiding developers to identify the use of software design patterns.

In order to repair a misapplied design pattern, we firstly need to identify design patterns and be able to differentiate between their correct and incorrect implementations. Identifying design pattern instances allows developers to garner deeper understanding of the underlying software system's structure and function. This is useful because it helps developers comprehend the architecture, recognise the intent behind code structures, and facilitates tasks such as debugging, refactoring, and extending the system without introducing errors \cite{shi2005reverse}. Experienced software developers can understand the roles of specific parts of a software system by analysing whether design patterns have been used or not \cite{gamma1995design}. It also helps to observe whether there are potential flaws in the design and implementation which will need to be rectified. However, identifying design patterns are often the domain of experienced developers, making it difficult for new developers to detect the implementations of such patterns in software code. Automating this process can ease this burden and also significantly reduce the time in understanding the code, thus aiding extensibility \cite{tsantalis2006design}. Experienced developers can use automated approaches to quickly get to parts of code that may need potential improvements. Additionally, automated identification of patterns, would make conversations within software teams easier, where team members can converse using identified pattern names at a high level, rather than low level code details. 

Detecting design patterns is challenging for several reasons. Firstly, a design pattern can be implemented in many ways depending on the use case, developer style, or programming language, making the same pattern appear differently across instances — especially when considering variants of patterns \cite{kouli2022feature}. Secondly, manual detection is time‑consuming and requires substantial effort to understand both the code and the architectural decisions, particularly as the size of the codebase increases, since relationships between classes, methods, and their interactions become more complex \cite{moreira2022based}. Thirdly, design patterns are often implicitly embedded in the code’s structure and behavior, rather than being explicitly visible as statements or snippets, requiring a deep semantic and behavioral analysis to uncover them \cite{wang2022research}.

Previous research on design‑pattern detection has primarily relied on static analysis techniques, which examine the structure, syntax, and dependencies of code without executing it \cite{tsantalis2006design}. While these methods can be effective, they often face limitations, such as being time consuming when dealing with large or complex codebases.

More recently, large language models (LLMs) have been explored for design‑pattern detection by reasoning about code semantically \cite{pan2025code,schindler2025llm,pandey2025design}. While LLMs show promise in detecting such patterns, prior studies show that the LLMs performance varies  with code complexity, contextual ambiguity, and implementation style.

 In this pilot study, we focus on five patterns, singleton, adapter, bridge, composite and decorator, and explore three different detection approaches across four LLM models -  Qwen2.5 Coder, Gemma 3, Nxcode-CQ and NextCoder, and two ensemble approaches. Our goal is to assess whether LLMs can complement or enhance traditional static‑analysis methods for design‑pattern detection, and whether the format in which we present the information matters. We pose the following two Research Questions (RQs) to guide our analysis:
\begin{itemize}
    \item \textbf{RQ1}: How effectively do different types of LLMs detect design patterns? {
    }
    \item \textbf{RQ2}: What is the impact of different input modalities (source code, PlantUML representations and text-based descriptions) on the performance of LLMs in detecting design patterns?{
    }
\end{itemize}
By systematically testing these approaches, we aim to understand both the limitations and the promise of AI‑driven code comprehension, and to identify under which conditions LLMs can effectively support developers in understanding, refactoring, and maintaining complex software systems.

\section{\uppercase{related work}}
\label{sec:related}

Automatic detection of design patterns has been extensively studied using traditional methods, including structural analysis using graph-based techniques, and machine learning. Recently, Large Language Models have been considered for studies as well.
In the two sub-sections below we provide an overview of traditional and LLM-based approaches for deisgn pattern detection.

\subsection{Traditional Approaches}

\textbf{Graph-based approaches:} 
One of the early works \cite{tsantalis2006design} proposed a method that represents both system structures and pattern specifications as matrices and iteratively computes similarity scores to detect matches, including modified variants that deviate from canonical forms. To handle large systems efficiently, the method partitions software into smaller subsystems based on inheritance hierarchies. Evaluations on JHotDraw, JRefactory, and JUnit demonstrated high precision with no false positives, highlighting the effectiveness of graph-based similarity techniques for scalable pattern detection.

Another work \cite{mayvan2017design} extended graph-theoretic methods by representing both software systems and target patterns as semantic graphs, framing detection as a graph matching problem. Large system graphs are partitioned into subgraphs to improve efficiency. 
Behavioral signature analysis further refines matches, distinguishing patterns with similar structures. The approach is language-independent, supports patterns with any number of roles, and handles variant implementations. Evaluations on JUnit, JHotDraw, and JRefactory demonstrated high precision and recall, outperforming other techniques and highlighting the continued evolution of graph-based strategies.

Despite the effectiveness of graph-based approaches in achieving good results for design pattern detection, a few inherent limitations constrain their practical applicability and scalability. Graph-based methods require explicit structural formalization of both target patterns and source code, which introduces substantial overhead in representing complex inter-class relationships. Furthermore, design pattern instances having structural variations or alternative implementations which cannot be captured by predefined pattern specifications introduces flexibility constraints.


\textbf{Machine Learning approaches:} Researchers have explored machine learning approaches that begins with code metric (i.e., feature) extraction from projects, followed by dimensionality reduction using principal component analysis and classification via angular distance \cite{chaturvedi2018design}. Pattern-related characteristics extracted from code are mapped to numerical values that are used as features of the models. Evaluation across linear, polynomial, support vector regression, and neural networks showed that support vector regression provided the most reliable detection performance. This work illustrates how data-driven ML techniques can complement structural methods such as graph-based methods, enabling automated detection in larger and more complex codebases. 

Another work \cite{alhusain2013towards} introduced another ML-based method that aggregates outputs from multiple existing pattern recognition tools to construct a training dataset, applied feature selection to build informative vectors, and trained artificial neural networks to classify pattern instances. Their evaluation on real-world Java code showed that ML can effectively handle the flexibility inherent in design pattern implementations, providing a foundation for tool-independent detection. 

Similarly, feature-based approaches like DPD$_{F}$ \cite{nazar2022feature} leverage semantic representations of code, including call graph information, embedded via Word2Vec. A supervised classifier trained on this representation achieved over 80\% precision and nearly 79\% recall, outperforming previous detectors such as FeatureMaps \cite{thaller2019feature} and MARPLE DPD \cite{zanoni2015applying}, while demonstrating practical runtime performance for real-world systems.

Despite the flexibility and scalability advantages of machine learning approaches for design pattern detection, a few critical limitations undermine their effectiveness. Firstly, ML-based methods inherently depend on the quality and representativeness of the training data - models trained on limited or biased datasets may fail to generalize to unseen codebases with different coding styles. Additionally, feature extraction remains a significant bottleneck, requiring careful manual engineering of code metrics that may not capture the semantic subtleties essential for detecting patterns implemented through unconventional structural arrangements.

\subsection{LLM-based Approaches}
More recently, large language models have emerged as a novel approach for design pattern detection, leveraging semantic reasoning when compared to traditional approaches. 

Abdeljalil et al. have proposed detecting Gang of Four (GoF) design patterns directly from PlantUML representations using LLMs \cite{abdeljalil2025use}. Code is reverse-engineered from the P-MART repository into PlantUML representations and analysed via prompt-driven LLM inference. While somewhat effective in recognising patterns, the method faces challenges when multiple patterns coexist, though adding lightweight comments to the PlantUML codes improves detection accuracy.

Another work investigated LLMs for design pattern detection, emphasizing their ability to reason about code semantics \cite{schindler2025llm}. Their study demonstrates that LLMs can identify many pattern instances without explicit structural analysis, but they struggle with nuanced architectural relationships and exhibit sensitivity to training data biases. This work highlights both the promise and limitations of LLMs as a stand-alone detection approach.

Pandey et al. conducted an empirical evaluation of multiple LLMs in detecting design patterns on Java code from the P-MART repository \cite{pandey2025design}. Using embeddings from models such as Code2Vec, CodeBERT, CodeGPT, CodeT5, and RoBERTa with a k-nearest neighbors classifier, they found that RoBERTa achieved the highest F1 score (~0.91), followed by CodeGPT and CodeBERT. Factors such as file size, coding style, and syntactic complexity influenced detection accuracy, illustrating that LLMs can capture pattern-relevant semantics even without explicit code pretraining.

Another work further examined whether Code LLMs can truly understand and generate design patterns in Python and Java \cite{pan2025code}. Their findings highlight frequent misclassification of common patterns like singleton and factory, especially in complex or abstract cases. Tasks involving design pattern classification, line completion, and function generation revealed significant biases and inconsistencies, emphasizing the gap between general code generation and pattern-aware reasoning. 

While LLM-based studies show promise, none of the prior works investigated the ability of LLMs in detecting design patterns when provided with several different types of inputs based on representations of the source code (e.g., code, PlantUML and descriptions). Such insights would help our understanding on where LLMs may support the drive to automate the detection of design patterns to help bootstrap new developers’ understanding of unfamiliar software system architectures, and support identifying and rectifying quality issues. This preliminary work bridges this gap, by investigating the detection of five specific design patterns - singleton, adapter, bridge, decorator and composite, and testing them on four different LLMs. 



\section{\uppercase{methodology}}
\label{sec:method}

Our goal here is to assess to what extent the state of the art LLMs, can be used to detect five design patterns (singleton, adapter, bridge, composite and decorator), using three different input types within LLM prompts: 1) Source code,  2) PlantUML descriptions of code and 3) Text-based descriptions of code. Further details of the methodology are elaborated below.

\subsection{Dataset}
\label{subsec:dataset}
The projects used for this research was taken from the P-MART dataset that was collected and published by a previous work \cite{gueheneuc2007p} and has been widely used in various studies (\cite{zanoni2015applying} and \cite{thaller2019feature}). The dataset contains source code, along with annotations stating which files contain the corresponding design pattern instances. The annotations are provided as an XML file, with various design pattern instances mentioned as entity elements within the specific pattern elements.

For our preliminary investigation, we selected five design patterns: singleton, adapter, bridge, composite and decorator. We aimed to cover most of the structural design patterns, with varied complexity, as well as \textcolor{red}{one singleton pattern (a creational pattern)}, due to its simplistic nature (for comparison purposes). This experimental setup allows us to pilot the design pattern detection capabilities of the LLMs over differing complexities.

From the dataset, we extracted files containing implementations of the five selected design patterns, yielding a total of 144 files (65 adapter, 5 decorator, 58 composite, 4 bridge and 12 singleton). To balance the dataset, an equal number of randomly selected files that did not contain the corresponding pattern was added for each category. Different LLMs were first evaluated using the source code of these files and then, in a second iteration, using their corresponding PlantUML representations. The PlantUML representations were generated by using a PlantUML Parser tool \footnote{\href{https://github.com/shuzijun/plantuml-parser}{https://github.com/shuzijun/plantuml-parser}}. For the final experiment, we generated text-based descriptions of the source codes in the files using the Qwen-3-Coder model (30B) which is not included in the experiments. This three-step process enables us to compare the models’ performance across code-based, PlantUML-based and description-based prompts.

\subsection{LLM Models}
\label{subsec:models}
We selected three state of the art code models and one base non-coder model for our experiments. The coder models considered were Qwen2.5 32B Coder\footnote{\href{https://openrouter.ai/qwen/qwen-2.5-coder-32b-instruct}{https://openrouter.ai/qwen/qwen-2.5-coder-32b-instruct:free}}, NextCoder-7B \cite{aggarwal2025robust} \footnote{\href{https://huggingface.co/Mungert/NextCoder-7B-GGUF}{https://huggingface.co/Mungert/NextCoder-7B-GGUF}} and NXCode-CQ 7B orpo\footnote{ \href{https://huggingface.co/rjmalagon/Nxcode-CQ-7B-orpo-Q8_0-GGUF}{\url{https://huggingface.co/rjmalagon/Nxcode-CQ-7B-orpo-Q8_0-GGUF}}}. The models were chosen by taking into consideration their larger context window size as well as good performance based on the Big Code Models Leaderboard\footnote{\href{https://huggingface.co/spaces/bigcode/bigcode-models-leaderboard}{https://huggingface.co/spaces/bigcode/bigcode-models-leaderboard}}. The Qwen2.5 Coder model was accessed via API inference provided by Openrouter, whereas the other two models were inferenced on a local machine with an 11th generation Intel i5-Core CPU. To enable the models to run on a local machine, 8-bit quantised versions of the models were used. These provide minimum precision loss in weights, resulting in extremely low quality loss \cite{jin2024comprehensive}. This design enables us to investigate to what extent smaller coder models fare when compared against the large one (Qwen2.5 Coder). Additionally, a non-coder model Gemma 3 (27B) was considered to evaluate its performance compared to the coder models. This was accessed via Google's free API and was provided with the same prompts, differing slightly with additions of start and end tokens due to the inherent architecture of the model.

We also considered two ensemble approaches with a majority vote policy based on the outputs from three models out of the four. In one approach (called Ensemble 1), we consider solely the coder models (Qwen2.5 Coder, Nxcode-CQ and NextCoder) and in the second approach (called Ensemble 2), we considered the three models with the best performances (Nxcode-CQ, NextCoder and Gemma 3). These approaches were implemented to test whether the ensembles fare better than the individual models. As an example for our majority voting policy, if two or more models out of the three models chose to output ``yes'' for a particular prompt (i.e., a design pattern is present), then the ensemble output is considered to be a ``yes'' as well. This approach allows us to determine if combining the outputs from multiple models and taking the most popular decision can match or even outperform the individual models.
 

\subsection{Procedures}
\begin{figure}[!h]
    \centering
    \includegraphics[width=1\linewidth]{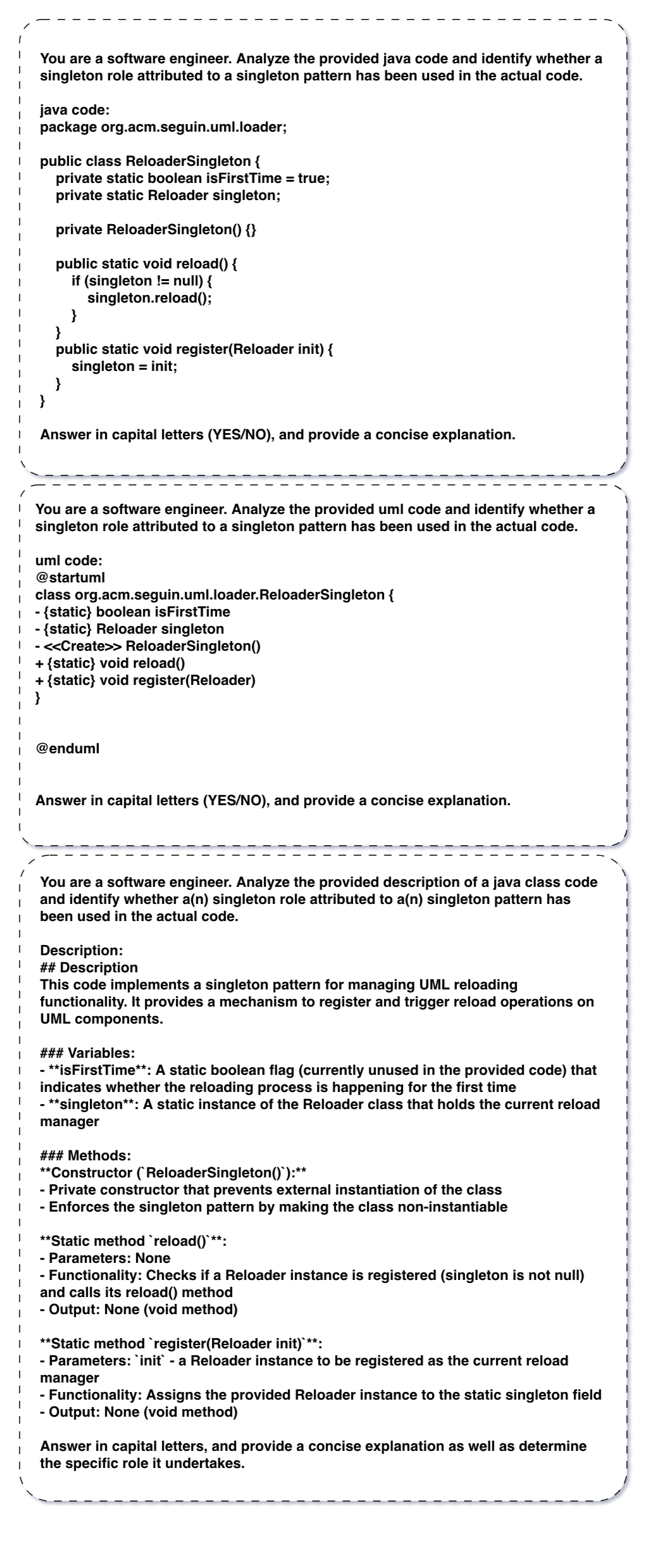}
    \caption{\centering Prompt snippets showing a) Code for singleton pattern (top), b) PlantUML code corresponding to the singleton pattern (middle) and c) Description corresponding to the same instance of singleton pattern (bottom)}
    \label{fig:prompts}
\end{figure}

For the three runs of the experiment the corresponding prompts contained a) Source code, b) PlantUML representation of the code and c) Descriptions of the methods and variables of the code (see Figure \ref{fig:prompts}), respectively. In all the cases, the models were asked to answer whether the specified design pattern role was used and the response collected from the models included a ``yes'' or a ``no'' along with an explanation for the answers they provided. These were then collated, analysed and the performance metrics (accuracy, precision, recall and F1 score) were tabulated. The results for the ensemble approach were also calculated from these results. Afterwards, appropriate statistical tests were performed to evaluate the significance of the reported results.

\section{\uppercase{Results}}
\label{sec:results}

This section presents the findings obtained from evaluating large language models on their ability to identify the singleton, adapter, decorator, bridge and composite design patterns based on three types of inputs (code, PlantUML and description). The performance of the models across the three different kinds of inputs are presented in Tables \ref{tab:all_codes}, \ref{tab:all_url} and \ref{tab:all_summary}. Additionally, graphs aggregating the performances of the models and the performances of the input modalities are presented in Figures \ref{fig:all_models} and \ref{fig:all_types}.


\begin{table}[t]
\centering

\footnotesize
\setlength{\tabcolsep}{4pt} 

\begin{tabularx}{\columnwidth}{
>{\centering\arraybackslash}X   
>{\centering\arraybackslash}X |   
*{6}{>{\centering\arraybackslash}X}  
}

\textbf{\rotatebox{90}{\parbox{2cm}{\centering Patterns}}} & \textbf{\rotatebox{90}{\parbox{2cm}{\centering Metrics}}} & \textbf{\rotatebox{90}{\parbox{2cm}{\centering Qwen 2.5 Coder}}} & \textbf{\rotatebox{90}{\parbox{2cm}{\centering Nxcode CQ}}} & \textbf{\rotatebox{90}{\parbox{2cm}{\centering NextCoder}}} & \textbf{\rotatebox{90}{\parbox{2cm}{\centering Gemma 3}}} & \textbf{\rotatebox{90}{\parbox{2cm}{\centering Ensemble 1}}} & \textbf{\rotatebox{90}{\parbox{2cm}{\centering Ensemble 2}}}\\
\hline

\multirow{4}{*}{\rotatebox{90}{Singleton}}
& Acc & 0.75 & 0.5 & \textbf{0.88} & \textbf{0.88} & 0.83 & \textbf{0.88} \\
& Prec & 0.8 & 0.5 & \textbf{0.91} & 0.85 & 0.83 & 0.85 \\
& Rec & 0.67 & \textbf{0.92} & 0.83 & \textbf{0.92} & 0.83 & \textbf{0.92} \\
& F1 & 0.73 & 0.65 & 0.87 & \textbf{0.88} & 0.83 & \textbf{0.88} \\
\hline

\multirow{4}{*}{\rotatebox{90}{Adapter}}
& Acc & 0.65 & 0.57 & 0.78 & 0.65 & 0.79 & \textbf{0.8} \\
& Prec & 0.88 & 0.54 & 0.84 & \textbf{0.92} & 0.85 & 0.85 \\
& Rec & 0.35 & \textbf{0.98} & 0.71 & 0.34 & 0.71 & 0.72 \\
& F1 & 0.51 & 0.7 & 0.77 & 0.49 & 0.77 & \textbf{0.78} \\
\hline

\multirow{4}{*}{\rotatebox{90}{Bridge}}
& Acc & 0.62 & 0.38 & \textbf{0.75} & 0.5 & \textbf{0.75} & 0.62 \\
& Prec & \textbf{1.0} & 0.4 & \textbf{1.0} & 0.5 & \textbf{1.0} & 0.67 \\
& Rec & 0.25 & \textbf{0.5} & \textbf{0.5} & 0.25 & \textbf{0.5} & \textbf{0.5} \\
& F1 & 0.4 & 0.44 & \textbf{0.67} & 0.33 & \textbf{0.67} & 0.57 \\
\hline

\multirow{4}{*}{\rotatebox{90}{Composite}}
& Acc & 0.55 & \textbf{0.59} & 0.53 & 0.53 & 0.53 & 0.57 \\
& Prec & \textbf{0.8} & 0.56 & 0.55 & 0.58 & 0.54 & 0.62 \\
& Rec & 0.14 & \textbf{0.9} & 0.31 & 0.19 & 0.33 & 0.36 \\
& F1 & 0.24 & \textbf{0.69} & 0.4 & 0.29 & 0.41 & 0.46 \\
\hline

\multirow{4}{*}{\rotatebox{90}{Decorator}}
& Acc & \textbf{1.0} & 0.6 & 0.9 & \textbf{1.0} & 0.9 & 0.9 \\
& Prec & \textbf{1.0} & 0.56 & 0.83 & \textbf{1.0} & 0.83 & 0.83 \\
& Rec & \textbf{1.0} & \textbf{1.0} & \textbf{1.0} & \textbf{1.0} & \textbf{1.0} & \textbf{1.0} \\
& F1 & \textbf{1.0} & 0.71 & 0.91 & \textbf{1.0} & 0.91 & 0.91 \\
\hline

\multirow{4}{*}{\rotatebox{90}{\makecell{Average \\ Performance}}}
& Acc & 0.71 & 0.53 & \textbf{0.77} & 0.71 & 0.76 & 0.75 \\
& Prec & \textbf{0.9} & 0.51 & 0.83 & 0.77 & 0.81 & 0.76 \\
& Rec & 0.48 & \textbf{0.86} & 0.67 & 0.54 & 0.67 & 0.7 \\
& F1 & 0.58 & 0.64 & \textbf{0.72} & 0.6 & \textbf{0.72} & \textbf{0.72} \\
\hline

\end{tabularx}
\caption{\centering Metrics for all models and patterns when codes have been evaluated.}
\label{tab:all_codes}
\end{table}

\begin{table}[t]
\centering

\footnotesize
\setlength{\tabcolsep}{4pt} 

\begin{tabularx}{\columnwidth}{
>{\centering\arraybackslash}X   
>{\centering\arraybackslash}X |   
*{6}{>{\centering\arraybackslash}X}  
}

\textbf{\rotatebox{90}{\parbox{2cm}{\centering Patterns}}} & \textbf{\rotatebox{90}{\parbox{2cm}{\centering Metrics}}} & \textbf{\rotatebox{90}{\parbox{2cm}{\centering Qwen 2.5 Coder}}} & \textbf{\rotatebox{90}{\parbox{2cm}{\centering Nxcode CQ}}} & \textbf{\rotatebox{90}{\parbox{2cm}{\centering NextCoder}}} & \textbf{\rotatebox{90}{\parbox{2cm}{\centering Gemma 3}}} & \textbf{\rotatebox{90}{\parbox{2cm}{\centering Ensemble 1}}} & \textbf{\rotatebox{90}{\parbox{2cm}{\centering Ensemble 2}}}\\
\hline

\multirow{4}{*}{\rotatebox{90}{Singleton}}
& Acc & 0.92 & 0.67 & 0.88 & \textbf{0.96} & \textbf{0.96} & 0.92 \\
& Prec & \textbf{1.0} & 0.61 & \textbf{1.0} & 0.92 & \textbf{1.0} & 0.92 \\
& Rec & 0.83 & 0.92 & 0.75 & \textbf{1.0} & 0.92 & 0.92 \\
& F1 & 0.91 & 0.73 & 0.86 & \textbf{0.96} & \textbf{0.96} & 0.92 \\
\hline

\multirow{4}{*}{\rotatebox{90}{Adapter}}
& Acc & 0.62 & 0.6 & 0.66 & \textbf{0.7} & 0.67 & 0.68 \\
& Prec & \textbf{1.0} & 0.56 & 0.84 & 0.96 & 0.84 & 0.83 \\
& Rec & 0.23 & \textbf{0.89} & 0.4 & 0.42 & 0.42 & 0.46 \\
& F1 & 0.38 & \textbf{0.69} & 0.54 & 0.58 & 0.56 & 0.59 \\
\hline

\multirow{4}{*}{\rotatebox{90}{Bridge}}
& Acc & \textbf{0.75} & 0.38 & \textbf{0.75} & \textbf{0.75} & \textbf{0.75} & \textbf{0.75} \\
& Prec & \textbf{1.0} & 0.33 & \textbf{1.0} & \textbf{1.0} & \textbf{1.0} & \textbf{1.0} \\
& Rec & \textbf{0.5} & 0.25 & \textbf{0.5} & \textbf{0.5} & \textbf{0.5} & \textbf{0.5} \\
& F1 & \textbf{0.67} & 0.29 & \textbf{0.67} & \textbf{0.67} & \textbf{0.67} & \textbf{0.67} \\
\hline

\multirow{4}{*}{\rotatebox{90}{Composite}}
& Acc & 0.54 & 0.56 & 0.62 & 0.59 & 0.6 & \textbf{0.65} \\
& Prec & \textbf{1.0} & 0.54 & 0.89 & 0.82 & 0.93 & 0.87 \\
& Rec & 0.09 & \textbf{0.74} & 0.28 & 0.24 & 0.22 & 0.34 \\
& F1 & 0.16 & \textbf{0.63} & 0.42 & 0.37 & 0.36 & 0.49 \\
\hline

\multirow{4}{*}{\rotatebox{90}{Decorator}}
& Acc & 0.9 & 0.7 & 0.9 & \textbf{1.0} & \textbf{1.0} & \textbf{1.0} \\
& Prec & \textbf{1.0} & 0.62 & \textbf{1.0} & \textbf{1.0} & \textbf{1.0} & \textbf{1.0} \\
& Rec & 0.8 & \textbf{1.0} & 0.8 & \textbf{1.0} & \textbf{1.0} & \textbf{1.0} \\
& F1 & 0.89 & 0.77 & 0.89 & \textbf{1.0} & \textbf{1.0} & \textbf{1.0} \\
\hline

\multirow{4}{*}{\rotatebox{90}{\makecell{Average \\ Performance}}}
& Acc & 0.75 & 0.58 & 0.76 & \textbf{0.8} & \textbf{0.8} & \textbf{0.8} \\
& Prec & \textbf{1.0} & 0.53 & 0.95 & 0.94 & 0.95 & 0.92 \\
& Rec & 0.49 & \textbf{0.76} & 0.55 & 0.63 & 0.61 & 0.64 \\
& F1 & 0.6 & 0.62 & 0.68 & 0.72 & 0.71 & \textbf{0.73} \\
\hline

\end{tabularx}
\caption{\centering Metrics for all models and patterns when PlantUML representations have been evaluated.}
\label{tab:all_url}
\end{table}

\begin{table}[t]
\centering

\footnotesize
\setlength{\tabcolsep}{4pt} 

\begin{tabularx}{\columnwidth}{
>{\centering\arraybackslash}X   
>{\centering\arraybackslash}X |   
*{6}{>{\centering\arraybackslash}X}  
}

\textbf{\rotatebox{90}{\parbox{2cm}{\centering Patterns}}} & \textbf{\rotatebox{90}{\parbox{2cm}{\centering Metrics}}} & \textbf{\rotatebox{90}{\parbox{2cm}{\centering Qwen 2.5 Coder}}} & \textbf{\rotatebox{90}{\parbox{2cm}{\centering Nxcode CQ}}} & \textbf{\rotatebox{90}{\parbox{2cm}{\centering NextCoder}}} & \textbf{\rotatebox{90}{\parbox{2cm}{\centering Gemma 3}}} & \textbf{\rotatebox{90}{\parbox{2cm}{\centering Ensemble 1}}} & \textbf{\rotatebox{90}{\parbox{2cm}{\centering Ensemble 2}}}\\
\hline

\multirow{4}{*}{\rotatebox{90}{Singleton}}
& Acc & 0.96 & 0.67 & 0.96 & 0.92 & \textbf{1.0} & 0.96 \\
& Prec & 0.92 & 0.62 & \textbf{1.0} & 0.92 & \textbf{1.0} & \textbf{1.0} \\
& Rec & \textbf{1.0} & 0.83 & 0.92 & 0.92 & \textbf{1.0} & 0.92 \\
& F1 & 0.96 & 0.71 & 0.96 & 0.92 & \textbf{1.0} & 0.96 \\
\hline

\multirow{4}{*}{\rotatebox{90}{Adapter}}
& Acc & 0.59 & 0.56 & \textbf{0.64} & 0.62 & 0.62 & 0.61 \\
& Prec & \textbf{0.88} & 0.56 & 0.6 & 0.6 & 0.62 & 0.59 \\
& Rec & 0.22 & 0.57 & \textbf{0.8} & 0.71 & 0.58 & 0.69 \\
& F1 & 0.35 & 0.56 & \textbf{0.69} & 0.65 & 0.6 & 0.64 \\
\hline

\multirow{4}{*}{\rotatebox{90}{Bridge}}
& Acc & 0.62 & 0.62 & \textbf{0.88} & \textbf{0.88} & 0.75 & \textbf{0.88} \\
& Prec & \textbf{1.0} & 0.6 & 0.8 & \textbf{1.0} & 0.75 & 0.8 \\
& Rec & 0.25 & 0.75 & \textbf{1.0} & 0.75 & 0.75 & \textbf{1.0} \\
& F1 & 0.4 & 0.67 & \textbf{0.89} & 0.86 & 0.75 & \textbf{0.89} \\
\hline

\multirow{4}{*}{\rotatebox{90}{Composite}}
& Acc & 0.57 & 0.57 & 0.54 & \textbf{0.66} & 0.58 & 0.64 \\
& Prec & \textbf{0.9} & 0.63 & 0.53 & 0.61 & 0.67 & 0.62 \\
& Rec & 0.16 & 0.33 & 0.81 & \textbf{0.84} & 0.31 & 0.72 \\
& F1 & 0.26 & 0.43 & 0.64 & \textbf{0.71} & 0.42 & 0.67 \\
\hline

\multirow{4}{*}{\rotatebox{90}{Decorator}}
& Acc & \textbf{1.0} & 0.4 & 0.5 & 0.8 & 0.5 & 0.5 \\
& Prec & \textbf{1.0} & 0.44 & 0.5 & 0.71 & 0.5 & 0.5 \\
& Rec & \textbf{1.0} & 0.8 & \textbf{1.0} & \textbf{1.0} & \textbf{1.0} & \textbf{1.0} \\
& F1 & \textbf{1.0} & 0.57 & 0.67 & 0.83 & 0.67 & 0.67 \\
\hline

\multirow{4}{*}{\rotatebox{90}{\makecell{Average \\ Performance}}}
& Acc & 0.75 & 0.56 & 0.7 & \textbf{0.78} & 0.69 & 0.72 \\
& Prec & \textbf{0.94} & 0.57 & 0.69 & 0.77 & 0.71 & 0.7 \\
& Rec & 0.53 & 0.66 & \textbf{0.91} & 0.84 & 0.73 & 0.87 \\
& F1 & 0.59 & 0.59 & 0.77 & \textbf{0.79} & 0.69 & 0.77 \\
\hline

\end{tabularx}
\caption{\centering Metrics for all models and patterns when text-based descriptions have been evaluated.}
\label{tab:all_summary}
\end{table}

\subsection{Comparison of LLM models (RQ1)}
Table \ref{tab:all_codes} shows the performance of the models aggregated over all the design patterns when provided with just the source code. To compare performances across models, we are using F1 scores, as it is more informative compared to just accuracy, and provides a balanced metric of the model's performance. However, we also report other metrics subsequently (e.g., precision and recall). The models perform relatively better in identifying the singleton and decorator patterns than the others, as can be seen from the relatively higher F1 scores (0.81 and 0.91 on average respectively). For the singleton pattern, Gemma 3 and NextCoder exhibit the best overall performances (0.88 and 0.87 respectively) with NextCoder achieving high precision and moderate recall, whereas Gemma 3 achieves a higher recall and moderate precision. Gemma 3 and Qwen 2.5 Coder perform the best in detecting the decorator pattern. Nxcode CQ performed the best in detecting the instances of the composite design pattern, achieving the highest F1 score of 0.69, with a high recall and low precision and accuracy. NextCoder outperformed the other models in detecting both the adapter and bridge design patterns, achieving the highest F1 score with high precision but lower recall, indicating a conservative prediction strategy where the model predicts that a pattern has been used only when it is absolutely sure, otherwise predicting ``no''. Both the ensemble approaches performed comparable to each other, as evidenced by their similar average F1 scores. Overall, both the ensemble approaches displayed promising results, with a performance almost as good as the models considered separately, if not better.

Table \ref{tab:all_url} reports the performance of the models aggregated over all the design patterns when provided with the PlantUML representation of the code containing the design pattern instance. The models can again be seen performing the best in singleton and detector patterns as displayed by the relatively higher F1 scores (0.89 and 0.92 respectively) achieved by all the models and ensemble approaches. In singleton pattern detection, Gemma 3 and Qwen2.5 Coder perform relatively better on their own (0.96 and 0.91 respectively), achieving high precision and recall leading to good F1 scores. Gemma 3 once again gets all predictions for the decorator design pattern correct. Following a similar trend as the code based prompts, Nxcode CQ outperforms all models in detecting the composite design patterns with the highest F1 score of  0.63. This model also performed very well in recognising the adapter pattern instances, once again achieving the highest F1 score of 0.69. However, it performed poorly for the bridge pattern, where all the other models outperformed Nxcode CQ. Of the 2 ensemble approaches considered, the second approach performed similar or better in most of the design patterns, only being outperformed in singleton pattern detection.

Table \ref{tab:all_summary} reports the performance of the models aggregated over all the design patterns when provided with the text-based descriptions of the elements of the source code. The models performed best in detecting the singleton instances once again, however a few of the models suffered minor dips in their performance when detecting the decorator design pattern when compared to other representations. NextCoder and Gemma 3 perform the best in detecting composite, bridge and adapter design patterns out of all the four models considered, indicated by their relatively higher F1 scores. Most models achieved high precision and recall in detecting the patterns, however, Qwen2.5 coder suffers from very low recall in adapter, bridge and composite patterns, indicating a very conservative prediction strategy followed by the model.

For all the representations of the design pattern instances, the performance varies considerably between the different models, particularly in recall and F1 scores. While several models achieve high precision, they also suffer from lower recall for most design pattern. The ensemble approaches generally demonstrate more balanced performance when compared to the individual models. This can be seen in the F1 scores, which indicate that aggregating the predictions from the model results in better performances.

\begin{figure*}[!ht]
    \centering
    \includegraphics[width = 0.9\linewidth]{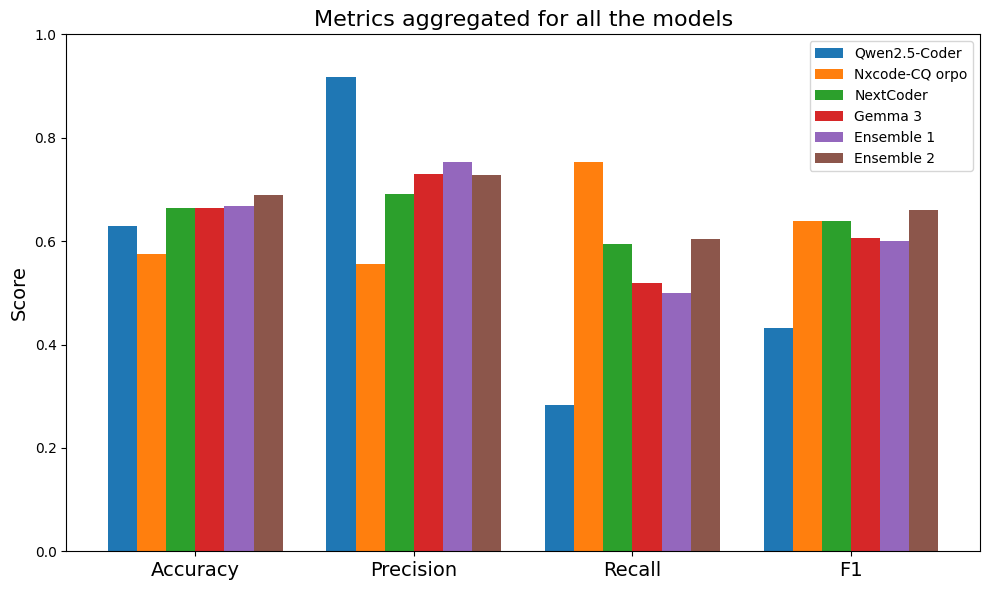}
    \caption{Comparison of the aggregated performance metrics of all models}
    \label{fig:all_models}
\end{figure*}

Figure \ref{fig:all_models} shows the performance metrics of all the models and ensemble approaches, aggregated over all the design patterns instances considered. NextCoder and Gemma 3 achieved the highest accuracies out of the four models (approx. 66\% for both), with relatively moderate precision scores (0.69 and 0.73 respectively), indicating some confidence in the positive classifications. However, they also suffer some of the lowest recall scores (0.59 and 0.52 respectively) pointing to a limited ability in identifying all the true instances. Comparatively, Qwen 2.5 Coder achieves the highest precision of 0.92, displaying strong confidence in the positive classifications made by the model. However, it suffers from the lowest recall score of 0.28, due to its inability in identifying true positives and consequently achieving the lowest F1 score of all the models. Of the two ensemble approaches, the second approach achieves a higher accuracy and recall, indicating a strong performance in finding positive instances of the design patterns, and resulting in achieving the highest F1 score compared to all other models. 

To analyse the results further, a Friedman test was performed on the F1 scores of the four models and two ensembles, which revealed a statistically significant difference among the 6 models, $\chi^2$(5) = 14.70, p $=$ 0.012. Follow-up post-hoc pairwise Conover test showed that Qwen 2.5 Coder performed significantly poorly when compared to Gemma 3 and both the ensembles (Ensemble 1 and Ensemble 2), with p $=$ 0.025, p $=$ 0.047 and p $<$ 0.002 respectively. Also, Nxcode-CQ performed significantly poorly when compared to Gemma 3, Ensemble 1 and Ensemble 2, with p $=$ 0.025, p $=$ 0.047 and p $<$ 0.002 respectively. An interesting observation from this test was that aggregating the outputs from all three coder models (i.e., Ensemble 1) resulted in statistically significant improvement over two of the three coder models individually.

\subsection{Comparison of three types of input modalities (RQ2)}
\begin{figure*}[!ht]
    \centering
    \includegraphics[width=0.9\linewidth]{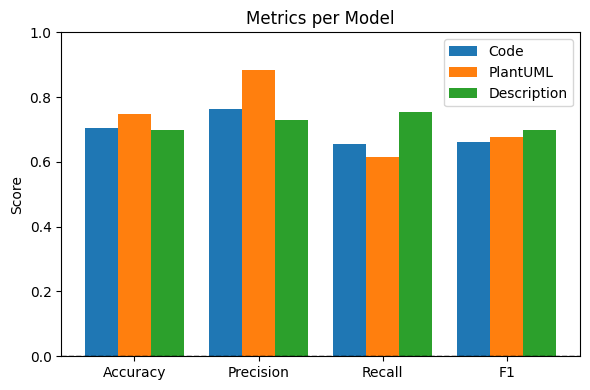}
    \caption{Comparison of the different types of representations of the source code.}
    \label{fig:all_types}
\end{figure*}

Figure \ref{fig:all_types} shows the performance metrics for the three types of representations of the design pattern instances considered (code, PlantUML and description), aggregated over all the models. All three representations achieve similar accuracy, but PlantUML representation achieves a much higher precision. Description-based prompts outperform the other two representations with the highest recall score of 0.75, indicating its better ability to present the necessary information for identifying more correct instances. Meanwhile, the PlantUML representations achieve high precision at the cost of a lower recall. Overall, all the representation types score similar in terms of F1 scores. As the F1 scores of the different types of representations of the pattern instances followed a normal distribution based on the Shapiro-Wilk test, a one-way repeated measures ANOVA test was performed which revealed that there is no significant difference in the representation approaches, with F(2, 58) $=$ 0.70 and p $=$ 0.50. This result shows that there is no statistical difference in how the design pattern information is presented to the models, further backing the results shown by the F1 scores in Figure \ref{fig:all_types}.

\section{\uppercase{Discussion}}
\label{sec:discussion}
Unlike prior studies, our study compared the performance of three coder and one non-coder LLM on five design patterns with three different prompting strategies (source code, PlantUML and text-based descriptions). We also looked at two ensemble approaches, one with all three coders and the other with the three best models. We note that ensemble models require extra time and computational overheads compared to using just one of the LLMs.

\subsection{Comparison of LLM models (RQ1)} Our results show that NextCoder, Nxcode-CQ and Gemma 3 can identify most of the design patterns relatively well across all design patterns as displayed by their comparable F1 scores in Figure \ref{fig:all_models}. Out of the four models, NextCoder, Gemma 3 and Qwen 2.5 Coder perform the best in identifying the singleton design pattern instances, whereas only NextCoder and Nxcode-CQ perform the best in adapter pattern detection. NextCoder also outperforms other models in identifying the bridge design pattern. However, its performance gets overshadowed by Gemma 3 and Qwen 2.5 Coder when detecting decorator patterns. Nxcode-CQ shows promising performance in identifying composite patterns compared to the rest of the models considered in this paper. Both the ensemble approaches perform comparably in detecting the singleton, decorator, adapter and bridge patterns. 

In summary, our findings addressing RQ1 indicate that ensemble strategies show meaningful promise for design pattern detection for code and PlantUML inputs. Ensemble 2 in particular shows more promise (see F1 score in Figure 2). However, when description of code is provided as input, the non-coder model Gemma 3 produces the best result (see average F1 score in Table 3). This highlights the promise of non-coder model for identifying design patterns from textual descriptions. Also, Gemma, a non-coder model, outperformed two out of the three other coder models (based on statistical testing), highlighting its utility for design pattern detection in general. We intend to investigate whether similar trends hold for other established non-coder models from OpenAI, Meta and other major AI developers. This brings into question, whether specialised coder models are needed for design pattern detection. 


\subsection{Comparison of three types of input modalities (RQ2)} Interestingly, our statistical comparison results pertaining to RQ2 showed no significant differences in performance across the three input modalities (code, PlantUML and description), which suggests that each modality alone allows for similar performance. This is surprising given the variability in the level of details available in each modality (e.g., PlantUML had a compact representation of the class and its relationships while description had detailed textual information). This is in contrast with prior work that shows PlantUML-based approaches yield somewhat lower performance for design pattern detection (accuracy ranging from 0.32 to 0.53 in the work of \cite{abdeljalil2025use}). Additionally, code description shows comparable performance against code and PlantUML (and in fact the raw results in Figure 3 show that F1 scores for code descriptions has the highest average). None of the prior works have investigated the code description as an input for design pattern detection, and thus our work adds a new input modality that can be considered, which is the domain where non-coder models may potentially excel as demonstrated by the Gemma 3 model results. Further, this opens up a new direction of future work where we will investigate a combination of the three modalities (i.e., source code, PlantUML representation and text-based descriptions of the class) to study whether providing all three modalities as inputs to a model improves the baseline results reported in this paper.


\subsection{Threats to validity} 
A threat to construct validity is the experimental set up for this position paper. Only single files were analysed instead of considering the multiple inter-connected files, due to limitations to the context length when utilizing smaller models. This could have impeded the performance for the composite pattern in particular, explaining the lowest scores it achieved.  Including the supporting classes such as clients, concrete implementations and dependency files could have led to better results for pattern detection, and this forms the basis of our future experimentation. A threat to external validity (generalization) is the number of design patterns considered in this experimentation (limited to five). Other GoF patterns will be considered in the future to gauge the capability of the models in recognising those design patterns. Also, we intend to  consider other design pattern datasets (other than PMART). In the future, the experiments could be made more nuanced by prompting the model to deduce the exact pattern and role that have been used in a provided class, instead of asking whether a particular pattern had been used. 

Additionally, the only prompting strategy considered in the current work is zero-shot prompting, where the model does not receive any prior examples or demonstrations of the task. Multiple enhancements can be made to the prompting strategies. One such enhancement that has been already planned is providing examples of the design patterns at hand, leading to one-shot or few-shot prompting. Another would be the investigation of the chain-of-thought process, which would allow the model to explicitly articulate its reasoning steps, such as identifying key relationships, analysing code structure, and mapping components to pattern roles, before arriving at a final classification.

\section{Conclusion}
\label{sec:conclusion}
This work investigated the competency of Large Language Models in detecting instances of the singleton, adapter, bridge, composite and decorator software design patterns, with three input types: a) source code only, b) PlantUML representation of the source code only, and c) Text-based description of the source code only. Of the four code models considered (Qwen 2.5 Coder, NXCode-CQ, NextCoder and Gemma 3), NextCoder and Gemma 3 show the most promise as evidenced by the results and statistical tests. Furthermore, utilising an ensemble approach improves overall detection accuracy, however it requires extra computational efforts. Additionally, the nature of the input modality considered (code, PlantUML and description of code) yield similar detection accuracies (i.e., there is no statistically significant difference across these modalities), suggesting researchers can use any of the three modalities for design pattern detection. While LLMs show promise in recognising the implementation of design patterns, there is still scope for future improvements and further study.

\bibliographystyle{apalike}
{\small
\bibliography{example}}



\end{document}